\font\titlefont=cmbx10 scaled\magstep1 \nopagenumbers
\magnification=\magstep1
\def\>{\rangle}
\def\<{\langle}
\def\e{{\rm e}}
\def\d{{\rm d}}

\def\veck{{\bf k}}
\def\vecn{{\bf n}}
\def\vecS{{\bf S}}
\def\vecm{{\bf m}}
\def\vecN{{\bf N}}
\def\vecM{{\bf M}}
\def\vecOmega{{\bf \Omega}}
\def\a{\alpha}

\def\h{\eta}
\def\co{\cos{\beta\over 2}}
\def\so{\sin{\beta\over 2}}
\def\ct{\cos{\tilde{\theta}\over2}}
\def\st{\sin{\tilde{\theta}\over 2}}
\def\c{\cos{{\theta}\over2}}
\def\s{\sin{{\theta}\over 2}}
\def\cl{\cos{\theta'\over2}}
\def\sl{\sin{\theta'\over2}}
\centerline{\titlefont The Heisenberg antiferromagnet:}
\smallskip
\centerline{\titlefont an explicitly rotational invariant formulation}
\bigskip
\centerline{J. da Provid\^encia, C. Provid\^encia, M. Brajczewska,}

\centerline{Departamento de F\'\i sica, Universidade de Coimbra}

\centerline{3000 Coimbra Portugal  }
\centerline{and}
\centerline{J. da Provid\^encia, Jr}

\centerline{Departamento de F\'\i sica, Universidade da Beira Interior}

\centerline{Covilh\~a, Portugal}
\bigskip
\centerline{\bf Abstract}
\medskip
A simple derivation of an explicitly  rotation invariant  Lagrangian
describing the dynamics of an antiferromagnetic spin system
is presented.
The scope of the derived Lagrangian is analysed in the context
of schematic models.
It is shown that the Lagrangian
describes the behaviour 
spin systems
from the anti-ferromagnetic to the ferromagnetic regimes.
\bigskip
\centerline{\bf 1. Introduction}
\bigskip
The theory of the two-dimensional Heisenberg antiferromagnet has been
attracting much attention over the past several years in connection
with the the problem of spin fluctuations in copper oxides.
For a review we refer the reader to the work of Manousakis [1].
In ref. [2], a Lagrangian  for the antiferromagnetic spin system
has been proposed which is explicitly invariant under rotations.
Following the ideas of ref. [2], a new and simpler derivation of such a
rotationally invariant Lagrangian is presented. 
Ideal systems made up of a small number of spins (two and four)
are investigated in the framework of the  derived Lagrangian.
It is shown that the Lagrangian
describes the behaviour these spin clusters
from the anti-ferromagnetic to the ferromagnetic regimes.
\bigskip
\centerline{\bf 2. General formalism}
\medskip
We consider a spin system described by the
Heisenberg Hamiltonian
$$\hat{H}_{Hei} = {J\over 2}\sum_l\sum_{l'\in\<l\>}\hat{\bf S}_{l} \cdot
\hat{\bf S}_{l'}, \ \ \ \hat{\bf S}_{l}\cdot\hat{\bf S}_{l} =
s(s+1),\eqno{(1)}
$$
where $\hat{\bf S}_{l}$ are spin operators, $\<l\>$ stands for the
set of the nearest neighbors of the site $l$, the index $l$ runs
over the sites of a two-dimensional square lattice,
$J>0$ is the exchange
constant 
corresponding to the anti-ferromagnetic (AF) spin
interaction, and $s$ is the magnitude of spin.
In the study of spin systems, it is
convenient to exploit the
representation of the grand partition function (GPF) 
or the generating functional of the
spin Green functions in the form of  functional integrals over
spin coherent states $|z\rangle$ 
due to Klauder [3,4] or over spin coherent states parametrized by
a unit vector
${\bf n}, \ {\bf n}^2=1,$ 
proposed by Manousakis [1]. These approaches are equivalent. Here,
we follow mostly the second alternative on account of the
simplicity of the corresponding measure of integration. We have,
$$\eqalignno{&Z = Tr\left[\exp\left(-\beta\hat{H}\right)\right], \ \ \
\beta = 1/T,\cr
&Z = \int_{-\infty}^{\infty} \cdots \int_{-\infty}^{\infty}
D\mu({\bf n}) \exp(A({\bf n})),&(2)
\cr  
&D\mu({\bf n})=\prod_{\tau l} {2s+1\over 2\pi} \delta({\bf
n}_{\tau l}^2-1)\d{\bf n}_{\tau l} }$$
where $T$ is the temperature, $\tau$ is the imaginary time, and
$A({\bf n})$ is the action of the system. In the continuum
approximation, which is valid in leading order in $1/2s$, the
expression of the action $A({\bf n}) $ simplifies and reads [3]
$$\eqalignno{ 
&A({\bf n}) = -\int_{0}^{\beta}\sum_{l}{\cal
L}_{tot}(\tau,l)\d\tau, \ \ \ {\cal L}_{tot}(\tau,l)={\cal
L}_{kin}(\tau,l)+{\cal H}(\tau,l),
\cr 
&{\cal L}_{kin}(\tau,l) = is(1-\cos\theta_{\tau l})
\dot{\phi}_{\tau l}, \ \ \ {\cal H}(\tau,l) = {Js^2\over
2}\sum_{l'\in\<l\>}{\bf n}_{\tau l} \cdot{\bf n}_{\tau l'},&(3)
}$$ 
where  $\theta, \phi$ are the Euler angles of the unit vector
${\bf
n}=(\cos\phi\sin\theta,\sin\phi\sin\theta,\cos\theta)$, and
${\bf \dot{\bf \phi}}_{\tau l}$ is the time derivative of
${\bf {\bf \phi}}_{\tau l}$.
The kinetic part of the 
Lagrangian density ${\cal L}_{kin}$ is highly nonlinear and it is
not clear how to proceed with it consistently. Essential steps
were taken in this direction in [1,5,6,7]
but we believe  the issue is not definitely settled yet.

In this paper we use the idea of near AF order. Accordingly, we
split our square lattice into two AF sublattices ${\rm a}$ and
${\rm b}$. In the sublattice ${\rm a}$ the spins ${\bf S}$ are normally
directed along some axis ${\bf \Omega }$, in the sublattice ${\rm
b}$ they are normally directed in the opposite direction. In this way, we
obtain a new square lattice with two spins ${\rm a}$ and ${\rm b}$
in the elementary cell with a volume $2a^2$, where $a$ is the
space distance between neighbouring spins. The axes of this new lattice are
rotated by 45 degrees with respect to the primary axes. We assume
that this AF order is only defined locally and any global AF order
is absent. Thus, the summations over the lattice sites $l$ and
$l'$ can be expressed as  summations over $l \in {\rm a}$ and $l'
\in {\rm b}$ which specifies the space positions of the spins in
the sublattices ${\rm a}$ and ${\rm b}$. The Lagrangian density
${\cal L}_{kin}$ is expressed as a sum of two Lagrangian
densities, one for the sublattice ${\rm a}$ and the other for the
sublattice ${\rm b}$, involving the vectors ${\bf n}_a(\tau,l)$
and ${\bf n}_b(\tau,l')$, respectively.  The Hamiltonian ${\cal
H}$ retains its form when the restrictions $l \in {\rm a}$ and
$l'\in {\rm b}$ are imposed but the multiplier $J/2$ must be
replaced by $J$ because the double summation disappears. In this
way we have two spins in each AF elementary cell which are defined
in different space positions $l$ and $l'$. Since we are assuming $J>0$,
spins in sublattice ${\rm a}$ tend to align themselves in the opposite
direction to the spins in the sublattice ${\rm b}$, for low $T$. However,
magnetic order is not, in principle, excluded. It would occur for $T>0,$ if $|T|$
is small enough, a condition which, physically, cannot be easily implemented.
\bigskip\noindent
\centerline{\bf 3. Coherent states}
\bigskip
Coherent states
may be defined as  
$$\eqalignno{|\vecn\rangle_s&=\e^{i\phi S_z}\e^{i\theta S_y}
\e^{-i\phi S_z}|s,s\rangle,&(4)
}$$ where $|s,s\rangle$ is such that
$$S_+|s,s\rangle=0, \quad
S_z|s,s\rangle=s|s,s\rangle.\eqno{(5)}$$ The overlap between two coherent
states reads 
$$
\eqalignno{_s\langle\vecn|\vecn'\rangle_s&=\left(\cos(\theta/2)\cos(\theta'/2)+
\sin(\theta/2)\sin(\theta'/2)\e^{i(\phi-\phi')}\right)^{2s},\cr
&=\left({1+\veck\cdot(\vecn+{\vecn}')+\vecn\cdot{\vecn}'+i\veck\cdot(\vecn\times{\vecn}')
\over
1+\veck\cdot(\vecn+{\vecn}')+\vecn\cdot{\vecn}'-i\veck\cdot(\vecn\times{\vecn}')
}\right)^s\left({1+\vecn\cdot\vecn'\over 2}\right)^s,&(6) }$$
where $\veck$ is an arbitrary unit vector. This expression lacks
manifest rotational invariance since it involves the unit vector
$\veck$ which defines the quantization direction. In the continuum
limit, this leads to the kinetic Lagrangian
$${L}_{kin}=is{\veck\cdot(\vecn\times\dot{\vecn})
\over 1+\veck\cdot \vecn},\eqno{(7)}
$$
connected with our inforced coherent state dynamics.
For a system of two spins, manifest rotational invariance may be
implemented. To this end, we start by allowing for a time
dependent quantization direction.
We introduce the kets $
|\vecn'\rangle_s',\,| s,s\rangle',$ which are defined analogously
to the previously introduced kets  $|\vecn\rangle_s,$ 
$|s,s\rangle,$ but refer to an arbitrary quantization direction
which need not be fixed in time. More specifically, we write
$$\eqalignno{
&
|\vecn'\rangle_s'=\e^{i\phi' S_{z'}}\e^{i\theta'
S_{y'}}\e^{-i\phi' S_{z'}}|s,s\rangle'\cr
&|s,s\rangle'=V|s,s\rangle=\e^{i\alpha S_z}\e^{i\beta
S_y}\e^{-i\alpha S_z}|s,s\rangle\cr
 &\e^{i\phi' S_{z'}}\e^{i\theta' S_{y'}}\e^{-i\phi' S_{z'}}=V
\e^{i\phi' S_{z}}\e^{i\theta' S_{y}}\e^{-i\phi' S_{z}} V^{-1}\cr
&|\vecn'\rangle_s'=V\e^{i\phi' S_{z}}\e^{i\theta'
S_{y}}\e^{-i\phi' S_{z}}|s,s\rangle. &{(8)}}
$$
Notice that  
$S_{z'}=V S_{z}V^{-1}$, 
$S_{y'}=V S_{y}V^{-1}$, 
$S_{z}|s,s\rangle=s|s,s\rangle$ and
$S_{z'}|s,s\rangle'=s|s,s\rangle'$. In order to compute the
overlap we need the relation between $|\vecn'\rangle_s'$ and $|s,s\rangle.$
The required Euler angles
$\tilde{\phi},\tilde{\theta},\tilde{\psi}$ characterizing the
unitary operator $V\e^{i\phi' S_{z}}\e^{i\theta' S_{y}}\e^{-i\phi'
S_{z}}= \e^{i\tilde{\phi} S_{z}}\e^{i \tilde{\theta}
S_{y}}\e^{-i(\tilde{\phi}-\tilde{\psi}) S_{z}}$
are such that
$$\eqalign{
\ct\e^{i({\tilde{\psi}/2})}&=\cl\co-\sl\so\e^{i(\alpha-\phi')}\cr
\st\e^{i(\tilde{\phi}-({\tilde{\psi}/2}))}&=\co\sl\e^{i\phi'}+\cl\so\e^{i\alpha}.
}$$ so that
$$\eqalignno{|\vecn\rangle_s
&=\sqrt{(2s)!}\sum_{m=-s}^{+s}{(\c)^{s+m}(\s\e^{i\phi})^{s-m}
\over\sqrt{(s+m)!(s-m)!}}|s,m\rangle,&(9) \cr
|\vecn'\rangle_s'&=\e^{i
s\tilde{\psi}}\sqrt{(2s)!}\sum_{m=-s}^{+s}{(\ct)^{s+m}(\st\e^{i\tilde{\phi}})^{s-m}
\over\sqrt{(s+m)!(s-m)!}}|s,m\rangle,&(10) \cr
_s\langle\vecn|\vecn'\rangle_s'&=\e^{i
s\tilde{\psi}}\left(\c\ct+\s\st\e^{i(\tilde{\phi}-\phi)}\right)^{2s}\cr
&=\e^{i s\tilde{\psi}}
\left({1+\veck\cdot(\vecn+{\vecn}')+\vecn\cdot{\vecn}'+i\veck
\cdot(\vecn\times{\vecn}') \over
1+\veck\cdot(\vecn+{\vecn}')+\vecn\cdot{\vecn}'-i\veck\cdot(\vecn\times{\vecn}')
}\right)^s\left({1+\vecn\cdot\vecn'\over 2}\right)^s ,&(11)\cr
}$$ where $\e^{i s\tilde{\psi}}$ is given by 
$$\eqalignno{
\e^{i\tilde{\psi}}&={\co\cl-\so\sl\e^{i
(\alpha-\phi')}\over\co\cl-\so\sl\e^{-i (\alpha-\phi')}},\cr
&={1+\veck\cdot(\veck'+{\vecn}')+{\veck}'\cdot{\vecn}'
+i\veck\cdot({\vecn}'\times{\veck}')\over
{1+\veck\cdot(\veck'+{\vecn}')+{\veck}'\cdot{\vecn}'
-i\veck\cdot({\vecn}'\times{\veck}')}}.&(12) }$$ In the continuum
limit, this leads to the kinetic Lagrangian
$${L}_{kin}=is{\veck\cdot(\vecn\times\dot{\vecn})
-\vecn\cdot(\veck\times\dot{\veck}) \over 1+\veck\cdot
\vecn}.\eqno(13)
$$
\noindent \centerline{\bf 4. A simple example }
\medskip
In order to illustrate the meaning of eq. (13), it is
convenient to consider the real time classical Lagragian for the
unit vector $\vecn$ parallel to the spin $\vecS$,
$${L}={\veck\cdot(\vecn\times\dot{\vecn})\over 1+\veck\cdot \vecn}-
{\cal H}(\vecn)-\lambda ((\vecn)^2-1) \eqno{(14)}
$$
where $\lambda$ is a Lagrange multiplier. The equation of motion
$$\dot{\vecn}=\vecn\times{\partial{\cal H}\over\partial \vecn}\eqno{(15)}
$$
does not depend on $\veck.$ If this vector 
changes with time,  the Lagrangian must be modified accordingly. Then,
the Lagrangian becomes
$${L}={\veck\cdot(\vecn\times\dot{\vecn})-\vecn\cdot(\veck\times\dot{\veck})
\over 1+\veck\cdot \vecn}-{\cal H}(\vecn)-\lambda ((\vecn)^2-1),\eqno{(16)}
$$
but the equation of motion remains unchanged.
\bigskip
\noindent\centerline{\bf 5. System of two interacting spins}
\bigskip
We discuss now the system of two interacting spins. We denote by
 $S_{cj},\,j=1,2,3,$ $ c=a,b$ 
the hermitian generators of $su(2),$ $[S_{c1},S_{c2}] = iS_{c3},$
etc., with $S_{c1}^2+S_{c2}^2+S_{c3}^2=s(s+1).$ We consider the
hamiltonian operator
$${\cal H}=J \vecS_a\cdot\vecS_b.\eqno{(17)}
$$
and investigate the collective modes, free energy, etc.
\medskip
Let $\vecS_t=\vecS_a+\vecS_b$. Then $s_t=0,1,\cdots,2s$ is the
quantum number characterizing the magnitude of the total spin and
$\vecS_a\cdot\vecS_b={1\over2}(\vecS_t\cdot\vecS_t-2s(s+1)).$ The
spectrum is
$$\eqalignno{&J
\left\{0-s(s+1),{1\over2}2-s(s+1),{1\over2}6-s(s+1),\cdots,\right.\cr
&\left.{1\over2}s_t(s_t+1)-s(s+1) ,\cdots,
{1\over2}2s(2s+1)-s(s+1)\right\}.&(18)}$$
The respective multiplicities
are
$$\{1,3,5,\cdots,2s_t+1,\cdots,4s+1\}.\eqno{(19)}
$$
This is the spectrum of a rigid rotor with moment of inertia
${\cal I}=J^{-1}$ whose angular momentum is constrained to a
maximum value $2s$.
\medskip
We follow now a mean-field variational approach. Here, mean-field
should be understood as an enforced coherent state dynamics of
each individual spin of our 2 spins system, i.e., it is assumed
that each spin is described by a coherent state   $|z\rangle$
which essentially depends on time through the parameter $z=z(t)$.
Consider
$$|z_1,z_2\rangle=\e^{z_1S_{a+}+z_2S_{b+}}|0\rangle,\;S_{a-}|0\rangle=S_{b-}|0\rangle=0,\;
S_{a0}|0\rangle=S_{b0}|0\rangle=-s|0\rangle.
$$
Since $\vecS_a\cdot\vecS_b={1\over2}(S_{a+}S_{b-}+S_{a-}S_{b+})+S_{a0}S_{b0},
$ we have
$${\langle z_1,z_2|\vecS_a\cdot\vecS_b
|z_1,z_2\rangle
\over\langle
z_1,z_2|z_1,z_2\rangle}=s^2{2(z_1^*z_2+z_2^*z_1)+(z_2^*z_2-1)(z_1^*z_1-1)\over
(z_2^*z_2+1)(z_1^*z_1+1)},
$$
so that
$$-s^2\leq\langle\vecS_a\cdot\vecS_b\rangle\leq s^2.
$$
The lower bound, $-s^2,$ is attained for $z_2=-1/z_1^*,$ (anti-parallel spins) while the upper bound,
$s^2,$ is attained for $z_2=z_1$ (parallel spins).
The mean field-Lagrangian describing our system of two interacting
spins may be written
$${L}=s{\veck\cdot(\vecn_a\times\dot{\vecn_a})\over 1+\veck\cdot \vecn_a}
+s{\veck\cdot(\vecn_b\times\dot{\vecn_b}) \over 1+\veck\cdot
\vecn_b}-{J}s^2(\vecn_a\cdot\vecn_b)-\lambda_a((\vecn_a)^2-1)-\lambda_b
((\vecn_b)^2-1)\eqno{(20)}
$$
where $\lambda_a,\lambda_b$ are Lagrange multipliers. Here,
rotational invariance is not manifest. The equations of motion
lead to
$$\dot{\vecn_a}=(\vecn_a\times\vecn_b) sJ,\quad\dot{\vecn_b}=-(\vecn_a\times\vecn_b) sJ,\eqno{(21)}
$$
or, with $\vecn=\vecn_b-\vecn_a$, $\vecm={1\over
2}(\vecn_a+\vecn_b),$
$$\dot{\vecn}=2(\vecm\times\vecn) sJ,\quad \dot{\vecm}=0.\eqno{(22)}
$$
From these equations it is not apparent that $|\dot{\vecn}|$ is
constrained to a maximum value, $|\dot{\vecn}|\leq 2s$.
However, there  is no incompatibility between these equations and such a constraint.
Identifying, in Eq. (13), $\veck$ with $-\vecn_b$ and $\vecn $
with $\vecn_a$, the previous Lagrangian may be replaced by
$${L}=-s{\vecn_b\cdot(\vecn_a\times\dot{\vecn_a})
+\vecn_a\cdot(\vecn_b\times\dot{\vecn_b}) \over 1-\vecn_a\cdot
\vecn_b}-{J}s^2(\vecn_a\cdot\vecn_b)-\lambda_a((\vecn_a)^2-1)-\lambda_b
((\vecn_b)^2-1).\eqno{(23)}
$$
Here, rotational invariance is manifest. This Lagrangian is of
the same form as the one implied
by eq. (8) of ref. [2]. The equations of motion
do not change when (22) is replaced by (23).
It is convenient to 
introduce new variables
$${\vecn_a+\vecn_b\over 2}={\vecN},\quad
{\vecn_a-\vecn_b\over 2}=\vecOmega\sqrt{1-N^2},\eqno{(24)}.
$$
We find
$$\eqalignno{{\vecN\cdot(\vecOmega\times\dot{\vecOmega})}
=&{-\vecn_b\cdot(\vecn_a\times\dot{\vecn_a})- \vecn_a\cdot(\vecn_b\times
\dot{\vecn_b})\over2( 1-\vecn_a\cdot\vecn_b)}. &(25)}$$
In terms of the new variables the Lagrangian becomes
$${\cal L}=2s{\vecN\cdot(\vecOmega\times\dot{\vecOmega})}-
{J}s^2\left(2N^2-1\right)-\lambda(\Omega^2-1)-\mu(\vecOmega\cdot \vecN)
$$
or,
$${\cal L}=s{\vecM\cdot\dot{\vecOmega}}-
{J}s^2\left({1\over2}{M^2}-1\right)-\lambda({\vecOmega}^2-1)-\mu(\vecOmega\cdot \vecM),\eqno{(26)}
$$
where
$$\vecM=2{\vecN\times\vecOmega},\quad 0\leq M\leq 2.\eqno{(27)}
$$

In terms of the variables $\vecM,\,\vecOmega$, the path integral mesure of integration becomes
$$\prod_\tau{(2s+1)^2\over 8\pi^2}\delta(\Omega^2-1)
\delta(\vecOmega\cdot\vecM)\d\vecOmega\d\vecM.\eqno{(28)}
$$

We observe that eq.(26) is the classical Lagrangian of a rigid rotor with moment of
inertia ${\cal I}=J^{-1}$ whose angular momentum $s\vecM$ is
constrained to a maximum value $2s$. The equations of motion may
be written
$$\dot\vecM\cdot\vecM=0,\quad\dot{\vecOmega}-s{J}\vecM=0,
\quad\dot{\vecM}+s{J}M^2\vecOmega=0,\quad\dot\vecN=0,\quad\vecN=
{1\over2}\vecOmega\times\vecM.\eqno{(29)}
$$
The mean field frequencies so obtained are in agreement with the
exact quantal results for the excitation energies in the whole
range from the 
AF ($M=0$) to the ferromagnetic
($M=2$) regimes.
\bigskip
\noindent {\bf Classical partition function}
\medskip
 In eq.(26), ${\cal L}$  is the classical Lagrangian of a rigid
rotor with moment of inertia ${\cal I}=J^{-1}$ whose angular
momentum $s\vecM$ is constrained to a maximum value $2s$.
The corresponding classical partition function may be written

$$\eqalignno{Z&=s^2\int_{0\leq p^2_\theta+p^2_\phi/\sin^2\theta\leq4} \d\phi\d\theta\d p_\phi
\d p_\theta\e^{-\beta
Js^2[(p^2_\theta+p^2_\phi/\sin^2\theta)/2-1]}\cr &=s^2\int_{0\leq
p^2_\theta+p'^2\leq4} \d\phi\d\theta\d p' \sin\theta \d
p_\theta\e^{-\beta Js^2[(p^2_\theta+p'^2)/2-1]}\cr
&={8\pi^2s^2\over\beta Js^2}\left(\e^{\beta Js^2}-\e^{-\beta
Js^2}\right). &(30)}
$$
The classical  free
energy, average energy and entropy, are
$$\eqalignno{-\beta F_{cl}&=\log 16\pi^2s^2
+\log{\sinh\beta Js^2\over \beta Js^2},&(31)\cr
\beta E_{cl}&=1
-\beta Js^2\coth\beta Js^2,&(32)\cr S_{cl}&=\log
16\pi^2s^2+1+\log{\sinh\beta Js^2\over \beta Js^2}-\beta Js^2\coth
\beta Js^2.&(33)}$$
These expressions are in good agreement with
the corresponding quantal results even for not very large $s$.
Indeed, if $s$ is not too small,  the quantal partition function,
$$Z_{q}=\sum_{\sigma=0}^{2s} (2 \sigma+1) \e^{-\beta J({1\over2}\sigma(\sigma+1)-s(s+1))}
$$
is well approximated by the classical partition function (30).
\bigskip
\noindent {\bf Path integral computation of the partition
function}
\medskip
From (26), integration over the field $\vecM$ after returning
to complex time,
 leads to
$${\cal L}=
{1\over 2J}
\dot\vecOmega^2-\lambda({\vecOmega}^2-1). 
\eqno{(34)}
$$
This is the continuum form associated with
$$\eqalignno{
Z&=\int
{\prod_k}\d\vecOmega_k{\d\lambda_k}\exp\left[-\sum_k\left({1\over
2J\epsilon}(\vecOmega_k-\vecOmega_{k+1})^2
+\epsilon\lambda({\vecOmega_k}^2-1)-\epsilon Js^2\right)\right]\cr
&=K{\exp(\beta Js^2)\over\beta}.&(35)}$$
This result is valid for
$\beta>0.$ It involves the approximation of neglecting the
constraint $|\dot{\vecOmega|}=Js|\vecM|\leq 2Js.$
\bigskip
\noindent {\bf Generator Coordinate Method and quantal fluctuations}
\medskip
For
$$|z\rangle=\e^{z S_+}|s,-s\rangle,\eqno{(36)}$$
where $|s,-s\rangle$ is such that
$S_-|s,-s\rangle=0, \quad
S_z|s,-s\rangle=-s|s,-s\rangle,$ we have
$$\eqalignno{
&\langle z|z' \rangle=(1+z'z^*)^{2s},\quad \langle z|S_+|z'
\rangle={2sz^*\over 1+z'z^*}(1+z'z^*)^{2s},\cr &\langle z|S_-|z'
\rangle={2sz'\over 1+z'z^*}(1+z'z^*)^{2s},\quad \langle z|S_0|z'
\rangle=-s{1-z'z^*\over 1+z'z^*}(1+z'z^*)^{2s}.&(37)}$$ Having in
mind investigating the expectation values of the operator ${\bf
S}_a\cdot{\bf S}_b$, we notice that
$$\eqalignno{&{\langle z|{\bf S}|z' \rangle\cdot\langle -z^{*-1}|{\bf S}|-z'^{*-1} \rangle
\over\langle z|z' \rangle \langle-z^{*-1}|-z'^{*-1}\rangle}
=-s^2{(1+z^*z)(1+z'^*z')+(z^*-z'^*)(z-z')\over(1+z^*z')(1+z'^*z)},\cr
&\langle z|z' \rangle\langle -z^{*-1}|-z'^{*-1}
\rangle=(1+z'z^*)^{2s}(1+{1\over{z'^*z}})^{2s}.&(38)}$$
According
to the mean field approximation, the expectation values of the
operator ${\bf S}_a\cdot{\bf S}_b$ lie between $-s^2$ (value which
is reached in (38) for $z=z'$) and $s^2$. In order to correct the
lower bound, it is instructive to apply the Generator Coordinate
Method to this system. In the present context, this amounts to
diagonalizing the Hamiltonian $H$ in the subspace spanned by the
states $|z' \rangle\otimes|-z'^{*-1} \rangle$, which are such that
the spins of the coherent states $|z' \rangle$ and $|-z'^{*-1}
\rangle$ point in opposite directions. To this end, let us assume
that $z=\eta$ and $z'=\eta'$ are real and the coherent states
$|\eta\rangle,\,|\eta'\rangle,|-\eta^{-1}\rangle,\,|-\eta'^{-1}\rangle,$
are normalized. Then the overlaps may be approximated by
$$\eqalignno{{\cal H}(\tilde\eta,\tilde\eta')&={\langle\eta|\otimes
\langle-\eta^{-1}|H|\eta'\rangle \otimes|-\eta'^{-1}\rangle\over
\langle\eta|\otimes\langle-\eta^{-1}|I|\eta'\rangle\otimes|-\eta'^{-1}\rangle}
\approx
-Js^2\left(1+2{(\eta-\eta')^2\over(1+\bar\eta^2)^2}\right)\cr&\approx
-Js^2\left(1+2{(\tilde\eta-{\tilde\eta}')^2}\right)&(39)\cr {\cal
N}(\tilde\eta,\tilde\eta')&=\langle\eta|\otimes\langle-\eta^{-1}|I|\eta'\rangle
\otimes|-\eta'^{-1}\rangle\approx\,
\exp\left({-2s{(\eta-\eta')^2\over(1+\bar\eta^2)^2}}\right)\cr
&\approx\,
\exp\left({-2s{(\tilde\eta-\tilde\eta')^2}}\right)&(40)}$$
 where
$\bar\eta=(\eta+\eta')/2$ and $\tilde\eta=\int\d\eta/(1+\eta^2)$.
Minimizing the expectation value of $H$ for a state of the form
$$|\Psi\rangle=\int\d\tilde\eta'f(\tilde\eta')|\eta'\rangle
\otimes|-\eta'^{-1}\rangle\eqno{(41)}
$$
it is found that the ground state energy is
$$E_0={\int\d\tilde\eta'{\cal H}(\tilde\eta,\tilde\eta'){\cal N}
(\tilde\eta,\tilde\eta') \over\int\d\tilde\eta'{\cal
N}(\tilde\eta,\tilde\eta')}=-Js^2\left(1+{1\over
s}\right).\eqno{(42)}
$$
It is also instructive to apply the Generator Coordinate Method to
a linear combination of states of the form $|\eta
\rangle\otimes|\eta\rangle$, which are such that the spins of both
coherent states  point in the same  direction. This calculation
confirms that the maximum energy eigenvalue is $Js^2$.
\bigskip\noindent
\centerline{\bf 6. Extended system: AF regime}
\bigskip
We return to our 2D Heisenberg antiferromagnet. In terms of the variables
$\bf \Omega,$ $\bf M$, defined for each antiferromagnet cell,
the Lagrangian density (per AF elementary cell) reads
$$\eqalignno{
&{\cal L}_{\Omega M}={\cal L}_{kin}+{\cal H},\quad {\cal
L}_{kin}=is\dot{\bf \Omega}\cdot{\bf M}, \quad{\cal
H}=Js^2\sum_{l'\in\langle l\rangle}H_{ll'},\cr &{H}_{ll'}=-({\bf
\Omega}\cdot{\bf
\Omega}')\sqrt{1-{1\over4}M^2}\sqrt{1-{1\over4}M'^2}+{1\over4}({\bf
\Omega}\cdot{\bf \Omega}')({\bf M}\cdot{\bf M}')-{1\over4}({\bf
M}'\cdot{\bf\Omega}) ({\bf M}\cdot{\bf\Omega}')\cr
&+{1\over2}({\bf \Omega}\times{\bf \Omega}')\cdot\left({\bf
M}'\sqrt{1-{1\over4}M^2}+{\bf
M}\sqrt{1-{1\over4}M'^2}\right),&(43) }$$ where ${\bf \Omega}={\bf
\Omega}_{\tau l},$ ${\bf \Omega}'={\bf \Omega}_{\tau l'},$ ${\bf
M}={\bf M}_{\tau l},$ ${\bf M'}={\bf M}_{\tau l'}.$
Here,  calligraphic symbols are used in order to stress that
the Lagrangian and Hamiltonian densities, respectively
${\cal L}$ and ${\cal H}$, refer to a specific AF elementary cell $l$.The quantity
$1-({\bf\Omega}\cdot{\bf\Omega}')={1\over2}({\bf
\Omega}-{\bf\Omega}')^2$ is quadratic in $|{\bf
\Omega}-{\bf\Omega}'|$. Moreover, in the quadratic order in $|{\bf
\Omega}-{\bf\Omega}'|$ and $M$, we have $({\bf
M}'\cdot{\bf\Omega}) ({\bf M}\cdot{\bf\Omega}')\approx0,\;({\bf
\Omega}\times{\bf \Omega}')\cdot\left({\bf
M}'\sqrt{1-{1\over4}M^2}+{\bf
M}\sqrt{1-{1\over4}M'^2}\right)\approx0.$ Thus, the physically
interesting quadratic part of the Lagrangian density is
$${\cal L}^{(2)}=is\dot{\bf \Omega}\cdot{\bf M}
+Js^2\sum_{l'\in\langle l\rangle}\left[1-({\bf \Omega}\cdot{\bf
\Omega}')+{1\over2}({\bf M}\cdot{\bf M})\right].\eqno{(44)}
$$
\bigskip
\noindent
\centerline{\bf 7. Extended system: ferromagnetic regime}
\bigskip
In the ferromagnetic regime, $M=2$ so that $H_{ll'}$ reduces to
$$H_{ll'}={1\over4}({\bf
\Omega}\cdot{\bf \Omega}')({\bf M}\cdot{\bf M}')-{1\over4}({\bf
M}'\cdot{\bf\Omega}) ({\bf M}\cdot{\bf\Omega}')={\bf N}\cdot{\bf
N'}.\eqno{(45)}
$$
The Lagrangian density (per AF elementary  cell) becomes, in this regime,
$${\cal L}_{ferro}=2is\dot{\bf Q}\cdot{\bf N}
+Js^2\sum_{l'\in\langle l\rangle}({\bf N}\cdot{\bf
N'}),\eqno{(46)}
$$
where $\dot{\bf Q}={\bf N}\times\dot{\bf \Omega}$ and $N=1$. If $J>0$,
this regime only occurs for negative temperatures.
\bigskip\noindent
\centerline{\bf 8. System of four spins}
\bigskip
The system we wish to consider now is
a cluster of spins  constituted by two antiferromagnetic cells
such as is described by the Heisenberg Hamiltonian
$$H=J(\vecS_{a1}\cdot\vecS_{b1}+\vecS_{a2}\cdot\vecS_{b2}+
\vecS_{a1}\cdot\vecS_{b2}+\vecS_{a2}\cdot\vecS_{b1}).
\eqno{(47)}$$
The geometrical arrangement of the spins in the cluster is a ring.
Some of the formulae in this section are  special cases of those given in section 6
and are only explicitly presented here for the sake of clarity.
The  classical Lagrangian corresponding to the Hamiltonian (47) reads
$$\eqalignno{{ L}&=s{-\vecn_{b1}\cdot(\vecn_{a1}\times
\dot{\vecn}_{a1})-\vecn_{a1}\cdot(\vecn_{b1}\times
\dot{\vecn}_{b1})\over( 1-\vecn_{a1}\cdot\vecn_{b1})}
+s{-\vecn_{b2}\cdot(\vecn_{a2}\times\dot{\vecn}_{a2})-
\vecn_{a2}\cdot(\vecn_{b2}\times
\dot{\vecn}_{b2})\over( 1-\vecn_{a2}\cdot\vecn_{b2})}\cr
&-Js^2(\vecn_{a1}\cdot \vecn_{b1}+\vecn_{a2}\cdot \vecn_{b2}
+\vecn_{a1}\cdot \vecn_{b2}+\vecn_{a2}\cdot \vecn_{b1})
&(48)}$$
If to each antiferromagnetic cell we  associate
the variables $\vecN\,,\vecOmega$ defined by eq. (24)
we obtain
$$\eqalignno{&\vecn_a=\vecN+\vecOmega\sqrt{1-N^2}\,,
\quad \vecn_b=\vecN-\vecOmega\sqrt{1-N^2}.}$$
In terms of the new variables,
the Lagrangian becomes
$$\eqalignno{{ L}&=2s{\vecN_1\cdot(\vecOmega_1\times
\dot{\vecOmega_1})}+2s{\vecN_2\cdot(\vecOmega_2\times\dot{\vecOmega_2})}\cr&-
{J}s^2\left(2N_1^2+2N^2_2-2+2\vecN_1\cdot\vecN_2-2
\vecOmega_1\cdot\vecOmega_2\sqrt{1-N_1^2}\sqrt{1-N_2^2}\right)\,.
&(49)}$$ To this Lagrangian it is necessary to add the subsidiary
conditions
$$\Omega_1^2-1=\Omega_2^2-1=\vecOmega_1\cdot
\vecN_1=\vecOmega_2\cdot \vecN_2=0\,.
$$
For convenience, we have used here the variables $\vecN,\,\vecOmega,$
instead of the variables $\vecM,\,\vecOmega$ used in in (43) for an extended antiferomagnet.
However, it is clear that, for two AF elementary cells, (43) and (49) are equivalent.
In the AF regime, $N_1,N_2$ and
$|\vecOmega_1-\vecOmega_2|$ are small quantities which, in the
classical  limit, vanish for the ground state. The ground state
energy (AF regime) is $-4Js^2.$ On the other hand,
in the ferromagnetic regime, we have $N_1=N_2=1.$
The upper-state classical energy (ferromagnetic regime) is $4Js^2.$

Low energy excitations (for which $N_1,N_2$ and
$|\vecOmega_1-\vecOmega_2|$ are small) are described by the
quadratic Lagrangian
$$\eqalignno{{ L}^{(2)}&=2s{\vecN_1\cdot(\vecOmega_1\times\dot{\vecOmega_1})}
+2s{\vecN_2\cdot(\vecOmega_2\times\dot{\vecOmega_2})}\cr&-
{J}s^2\left(3N_1^2+3N^2_2+2\vecN_1\cdot\vecN_2+(\vecOmega_1-\vecOmega_2)^2
\right)&{(50)} }$$ It is convenient to consider the quantities
$$\vecM_1=2\vecN_1\times\vecOmega_1, \quad \vecM_2=2\vecN_2\times\vecOmega_2
$$
in terms of which we have
$$\eqalignno{{ L}^{(2)}&=s{\vecM_1\cdot\dot{\vecOmega}_1}+s{\vecM_2\cdot\dot{\vecOmega}_2}-
{J\over4}s^2\left(3M_1^2+3M^2_2+2\vecM_1\cdot\vecM_2+4(\vecOmega_1-\vecOmega_2)^2
\right)\cr&={s\over2}(\vecM_1+\vecM_2)\cdot
(\dot{\vecOmega}_1+\dot{\vecOmega}_2)+
{s\over2}(\vecM_1-\vecM_2)\cdot
(\dot{\vecOmega}_1-\dot{\vecOmega}_2)\cr&
-{J\over4}s^2\left(2(\vecM_1+\vecM_2)^2+(\vecM_1-\vecM_2)^2+4(\vecOmega_1-\vecOmega_2)^2
\right)\,.&{(51)}. }$$
 It follows that $\vecM_1+\vecM_2$
is a constant of motion. In order to present its interpretation we
go back to (49), making $\vecOmega_1=\vecOmega_2=\vecOmega, \,
\vecN_1=\vecN_2=\vecN$. This leads to the Lagrangian of a rigid rotor
$$\eqalignno{{ L}'&=4s{\vecN\cdot(\vecOmega\times\dot{\vecOmega})}
-{J}s^2\left(8N^2-4\right)\cr&
=s\vecM_t\cdot\dot\vecOmega-{J\over2}s^2(M_t^2-8).}$$
 Here, $\vecM_t=4\vecN\times\vecOmega$.

We investigate now the energy spectrum. The Hamiltonian
associated with the Lagrangian (51), namely $${\cal H}=
{J\over4}s^2\left(3M_1^2+3M^2_2+2\vecM_1\cdot\vecM_2+4(\vecOmega_1-\vecOmega_2)^2
\right),$$ is the sum
of a term describing a rigid rotor
$${J\over2}s^2(M_t^2-8)
$$
and another term describing a harmonic oscillator on a sphere,
$${J}s^2\left(\left({\vecM_1-\vecM_2\over2}\right)^2+(\vecOmega_1-\vecOmega_2)^2\right),
$$
which, for simplicity, we identify with a simple two dimensional
harmonic oscillator. This is, admittedly, a rather drastic
approximation. It is then natural to assume that the energy
spectrum is given by
$$E_{j,\nu_1,\nu_2}=J\left(-4s(s+1)+{1\over2}j(j+1)\right)+2Js(1+\nu_1+\nu_2).\eqno{(52)}
$$
with $j=0,1,\cdots, 4s$ and $\nu_1,\nu_2=0,1,2$.
 The multiplicity
associated with the quantum number $j$ is $2j+1.$
For $s={1\over 2},$ this expression predicts
 4 energy levels, namely $-2J\,,
-J\,,0\,,J\,.$

The eigenvalues and eigenvectors of the hamiltonian (47) are
easily obtained and are shown in the Appendix. It is remarkable
that  the exact energy spectrum of (47) coincides with the one
predicted by (52). Unfortunately, the interpretation of the
multiplicities is not so straightforward.
The multiplicities $1\,,3\,,7\,,5$,
determined by the exact diagonalization of the Hamiltonian (47),
may be understood as follows. The lowest energy level
$E=-2J$, corresponds to $j=\nu_1=\nu_2=0.$ The second energy
level, $E=-J$, corresponds either to $j=1$, or to $(\nu_1=1\,,
\nu_2=0)$ or to $(\nu_2=1\,, \nu_1=0).$ The
multiplicity  3 of this level may be explained assuming that
the oscillator level $(\nu_1=1\,, \nu_2=0)$ coincides with the
rotor level ($j=1\,,m=1)$ while the oscillator level $(\nu_1=0\,,
\nu_2=1)$ coincides with the rotor level $(j=1\,,m=-1)$ . The third energy level $E=0$,
corresponds either to the $j=1$ rotor levels combined with one of
the harmonic oscillator levels $(\nu_1=1\,, \nu_2=0)$, or
$(\nu_1=0\,, \nu_2=1)$, or to the rotor level $j=0$ combined with
$(\nu_1=2\,, \nu_2=0)$, or
 $(\nu_1=1\,, \nu_2=1)$, or
$(\nu_1=0\,, \nu_2=2)$.  The
multiplicity  7 of this level may be explained assuming that the
combination of
$(j=1\,,m=1)$ with $(\nu_1=1\,, \nu_2=0)$ is equivalent to the
combintion of $j=0$ with $(\nu_1=2\,,\nu_2=0)$ and  the
combination of $(j=1\,,m=-1)$ with $(\nu_1=0\,, \nu_2=2)$ is
equivalent to the combintion of $j=0$ with $(\nu_1=0\,,\nu_2=1).$
The fourth and highest energy
level $E=J$ is simply made up of the 5 rotational states $j=2$.
\bigskip
\centerline{\bf 9. Concluding remarks}
\medskip
A simple derivation of an explicitly  rotation invariant  Lagrangian
appropriate to a path integral description of the dynamics of an AF spin system
was presented. The scope of the derived Lagrangian was analysed in the context
of schematic models. In particular, it was found that the properties of clusters of
two and four spins are well accounted for by the derived Lagrangian.
It was shown that the Lagrangian obtained covers the whole region
from the AF to the ferromagnetic regimes.

\bigskip
\centerline{\bf Acknowledgement}
\medskip
Valuable critical remarks of one Referee are gratefully acknowledged.
This work was supported by the  Foundation for Science and Technology (Portugal)
through the the Project POCTI/ FIS/451/94.

\bigskip\noindent
\centerline{\bf References}
\bigskip
\item{[1]} E. Manousakis, Rev. Mod. Phys. 63, (1991), 1.
\item{[2]} V. I. Belinicher and J. da Provid\^encia, Ann. Phys. {\bf 298} (2002) 186.
\item{[3]} J.R. Klauder, and B.S. Skagerstam, Coherent States, World Scientific
Publishing, Singapore, 1985.
\item{[4]} V. R. Vieira and P. D. Sacramento, Ann. Phys. {\bf 242}(1995)188; Nucl. Phys. B {\bf 448 (1995) 331.}
\item{[5]} F. D. M. Haldane, Phys. Lett. A {\bf 93}(1983)464; Phys. Rev. Lett {\bf 50}(1983)1153;
Phys. Rev. Lett {\bf 61}(1988)1029.
\item{[6]} A. Auerbach, ``Interacting Electrons and Quantum Magnetism", Springer-Verlag, Berlin/New York, 1994.
\item{[7]} S.Sachdev, 
in ``Low Dimensional Quantum Field Theories for Condensed Matter Physicists"
edited by Y. Lu, S.Lundqvist, and G. Morandi, World Scientific,
Singapore, 1995.

\bigskip
\centerline{\bf Appendix}
\medskip
For completeness,
the eigenvalues and eigenvectors of the hamiltonian (47) are
presented in the Appendix. They may be written
\def\ua{\uparrow}
\def\da{\downarrow}
\def\ra{\rangle}
$$\eqalign{
E=-2J;\quad
&|0\ra=
{1\over{2\sqrt{3}}}(2|\ua\da\ua\da\ra+2|\da\ua\da\ua\ra-|\da\ua\ua\da\ra\,,
-|\ua\da\da\ua\ra-|\ua\ua\da\da\ra-|\da\da\ua\ua\ra)\,,\cr
E=-J;\quad
&|1,1\ra=
{1\over2}(|\ua\ua\ua\da\ra+|\ua\da\ua\ua\ra-|\ua\ua\da\ua\ra-|\da\ua\ua\ua\ra)\,,\cr
&|1,2\ra=
{1\over\sqrt{2}}(|\ua\da\ua\da\ra-|\da\ua\da\ua\ra)\,,\cr
&|1,3\ra=
{1\over2}(|\da\da\da\ua\ra+|\da\ua\da\da\ra-|\da\da\ua\da\ra-|\ua\da\da\da\ra)\,,\cr
E=0;\quad
&|2,1\ra=
{1\over2}(|\ua\ua\ua\da\ra-|\ua\da\ua\ua\ra-|\ua\ua\da\ua\ra+|\da\ua\ua\ua\ra)\,,\cr
&|2,2\ra=
{1\over2}(|\ua\ua\ua\da\ra-|\ua\da\ua\ua\ra+|\ua\ua\da\ua\ra-|\da\ua\ua\ua\ra)\,,\cr
&|2,3\ra=
{1\over2}(|\da\ua\ua\da\ra+|\da\da\ua\ua\ra-|\ua\ua\da\da\ra-|\ua\da\da\ua\ra)\,,\cr
&|2,4\ra=
{1\over2}(|\da\ua\ua\da\ra-|\da\da\ua\ua\ra+|\ua\ua\da\da\ra-|\ua\da\da\ua\ra)\,,\cr
&|2,5\ra=
{1\over2}(|\da\ua\ua\da\ra-|\da\da\ua\ua\ra-|\ua\ua\da\da\ra+|\ua\da\da\ua\ra)\,,\cr
&|2,6\ra=
{1\over2}(|\da\da\da\ua\ra-|\da\ua\da\da\ra-|\da\da\ua\da\ra+|\ua\da\da\da\ra)\,,\cr
&|2,7\ra=
{1\over2}(|\da\da\da\ua\ra-|\da\ua\da\da\ra+|\da\da\ua\da\ra-|\ua\da\da\da\ra)\,,\cr
E=J;\quad
&|3,1\ra=
|\ua\ua\ua\ua\ra\,,\cr
&|3,2\ra=
{1\over2}(|\ua\ua\ua\da\ra+|\ua\da\ua\ua\ra+|\ua\ua\da\ua\ra+|\da\ua\ua\ua\ra)\,,\cr
&|3,3\ra=
{1\over{\sqrt{6}}}(|\ua\da\ua\da\ra+|\da\ua\da\ua\ra+|\da\ua\ua\da\ra
+|\ua\da\da\ua\ra+|\ua\ua\da\da\ra+|\da\da\ua\ua\ra)\,,\cr
&|3,4\ra=
{1\over2}(|\da\da\da\ua\ra+|\da\ua\da\da\ra+|\da\da\ua\da\ra+|\ua\da\da\da\ra)\,,\cr
&|3,5\ra=
|\da\da\da\da\ra\,.}
$$
Here, $|0\ra$ denotes the ground state vector and $|i,j\ra$ denotes the state vector
of the $j$th member of the $i$th excited energy eigenspace. Otherwise an obvious notation
is used.
\end

\bibitem{Man1}
E. Manousakis, Rev. Mod. Phys. 63, (1991), 1.

\bibitem{Cha1}
S. Chakravarty, B. I. Halperin, and D. R. Nelson. Phys. Rev. B 39 (1989)
2344.

\bibitem{Sac1}
A. Chubukov, S. Sachdev, and J. Ye. Phys. Rev. B 49 (1994) 11919.

\bibitem{Pi1}
A. Sokol, and D. Pines. Phys. Rev. Lett. 71, (1993) 2813.

\bibitem{Pol1} A.M. Polyakov,
Gauge Fields and Strings, Harwood, New York (1987).

\bibitem{Zin1} J. Zinn--Justin, Quantum Field Theory and Critical
Phenomena, Oxford Science Publication, Clarendon Press, Oxford (1996).

\bibitem{Kla1}
J.R. Klauder, and B.S. Skagerstam, Coherent States, World Scientific
Publishing, Singapore, 1985.

\bibitem{Sac2} S.Sachdev, Quantum Antiferromanets in Two Dimensions,
in Low Dimensional Quantum Field Theories for Condensed Matter Physicists
edited by Y. Lu, S.Lundqvist, and G. Morandi, World Scientific,
Singapore, 1995.

\bibitem{Per1} A. Perelomov, {\it Generalized Coherent States and
Their Applications}, Springer Verlag, Berlin, 1986.

\bibitem{Var1} D.A. Varshalovich, A.N. Moskalev, V.K. Chersonskii,
Quantum theory of the angular momentum, Moscow, Nauka, 1975.

\bibitem{Bel1}  V.I. Belinicher, A.L. Chernyshev, L.V.Popovich, and
V.A.Shubin, Czech. Jour. of Phys., 1996, v. 46, Supl.

\medskip
\noindent {\bf Mean field, one-spin system}
\medskip
\def\tr{{\rm tr}}
Density matrix
$$D={\e^{-\gamma S_0}\over \tr\, \e^{-\gamma S_0}},\quad
\tr\,{\e^{-\gamma S_0}}={\sinh((s+{1\over2})\gamma)\over
\sinh({1\over2}\gamma)}
$$
$$\langle S_0\rangle={\tr S_0 D}=-(s+{1\over2})\coth((s+{1\over2})\gamma)+
{1\over2}\coth({1\over2}\gamma)
$$
$$\eqalign{{\cal S}&=-\tr\,D\,\log\,D\cr&=-(s+{1\over2})\gamma\coth((s+{1\over2})\gamma)+
{1\over2}\gamma\coth({1\over2}\gamma)+\log\,\sinh((s+{1\over2})\gamma)
-\log\,\sinh({1\over2}\gamma) }$$
\medskip
\noindent {\bf Mean field, two-spin system}
\medskip
It may be interesting to describe the two spin system by an
independent spin density matrix to see how important correlations
are. Consider the density matrix
$$D_{ab}={\e^{-\gamma (S_{a0}+S_{b0})}\over \tr_{ab}\, \e^{-\gamma (S_{a0}+S_{b0})}},\quad
\tr_{ab}\,\e^{-\gamma
(S_{a0}+S_{b0})}=\left({\sinh((s+{1\over2})\gamma)\over
\sinh({1\over2}\gamma)}\right)^2
$$
Expectation value of the hamiltonian $\vec S_{a}\cdot\vec S_{b},$
$$\langle \vec S_{a}\cdot\vec S_{b}\rangle={\tr_{ab} S_{a0}S_{b0} D_{ab}}=(-(s+{1\over2})\coth((s+{1\over2})\gamma)+
{1\over2}\coth({1\over2}\gamma))^2
$$
Entropy
$$\eqalign{{\cal S}_{ab}&=-\tr_{ab}\,D_{ab}\,\log\,D_{ab}\cr&=
2\left[-(s+{1\over2})\gamma\coth((s+{1\over2})\gamma)+
{1\over2}\gamma\coth({1\over2}\gamma)+\log\,\sinh((s+{1\over2})\gamma)
-\log\,\sinh({1\over2}\gamma)\right] }$$

\bigskip
\noindent {\bf Mean field, two-spin system}
\medskip
We may also consider the density matrix
$$D_{ab}={\e^{-\gamma (S_{a0}-S_{b0})}\over \tr_{ab}\, \e^{-\gamma (S_{a0}-S_{b0})}},\quad
\tr_{ab}\,\e^{-\gamma
(S_{a0}-S_{b0})}=\left({\sinh((s+{1\over2})\gamma)\over
\sinh({1\over2}\gamma)}\right)^2
$$
There is now a change in sign in the expectation value of the
hamiltonian $\vec S_{a}\cdot\vec S_{b},$
$$\langle \vec S_{a}\cdot\vec S_{b}\rangle={\tr_{ab} S_{a0}S_{b0} D_{ab}}=-(-(s+{1\over2})\coth((s+{1\over2})\gamma)+
{1\over2}\coth({1\over2}\gamma))^2
$$
However, the entropy remains unchanged.
$$\eqalign{{\cal S}_{ab}&=-\tr_{ab}\,D_{ab}\,\log\,D_{ab}\cr&=
2\left[-(s+{1\over2})\gamma\coth((s+{1\over2})\gamma)+
{1\over2}\gamma\coth({1\over2}\gamma)+\log\,\sinh((s+{1\over2})\gamma)
-\log\,\sinh({1\over2}\gamma)\right] }$$

It is now a simple matter to compute the (average value of) the
energy, the free energy, etc. as function of the temperature. We
have to determine $\gamma$ by minimizing (maximazing) the free
energy. Recall that $\beta$ may be positive or negative.
\bigskip
It turns out that "mean field" provides only a qualitative
decription for nonvanishing temperature. The classical description
is superior.

\noindent {\bf Heisenberg antiferromagnet}
\medskip
By the superscripts $(a),(b)$ we denote quantities referring,
respectively, to the sublattices $a,b.$ It is convenient to
introduce new variables
$$\vec n^{(a)}={\vec\Omega(1-(L^2/4))+\vec L\over 1+(L^2/4)},\quad
\vec n^{(b)}={-\vec\Omega(1-(L^2/4))+\vec L\over 1+(L^2/4)}.
$$
Then, we obtain
\def\na{n^{(a)}}
\def\nb{n^{(b)}}

$$
\eqalign{& {\na_x\dot \na_y-\na_y\dot\na_x\over 2(1-\na_z)}
 +{\nb_x\dot \nb_y-\nb_y\dot\nb_x\over 2(1-\nb_z)}
=\left[1+{3-\Omega_z^2\over(1-\Omega_z^2)^2}\Omega_z^2\right]L_z
(\Omega_y\dot\Omega_x-\Omega_x\dot\Omega_y)\cr
&+\left[1+{3-\Omega_z^2\over(1-\Omega_z^2)^2}\Omega_z^2\right]
\dot\Omega_z(\Omega_x L_y-\Omega_y L_x)
+\left[1-{3-\Omega_z^2\over 1-\Omega_z^2}\right]
\Omega_z(L_x\dot\Omega_y-L_y\dot\Omega_x)+{\d\over\d
t}F(\vec\Omega, \vec L)\cr &=(\vec
L\times\vec\Omega)\cdot\dot{\vec\Omega}
-{3-\Omega_z^2\over(1-\Omega_z^2)^2}\Omega_z
[(L_x\Omega_x+L_y\Omega_y)(\Omega_y\dot\Omega_x-\Omega_x\dot\Omega_y)\cr
&+(\Omega_x\dot\Omega_x+\Omega_y\dot\Omega_y)(\Omega_x
L_y-\Omega_y L_x)
+(\Omega_x^2+\Omega_y^2)(L_x\dot\Omega_y-L_y\dot\Omega_x)]\cr
&=(\vec L\times\vec\Omega)\cdot\dot{\vec\Omega}\,.}
$$
In the sequel, undashed variables are associated with sites in
sublattice $a$, while dashed variables are associated with nearest
neighbour sites in sublattice $b$,
$$\eqalign{\sum \vec n\cdot{\vec n}' &\approx\sum(\vec\Omega(1-L^2/2)+\vec L)
\cdot(-\vec\Omega'(1-L'^2/2)+\vec L')\cr
&\approx\sum(-\vec\Omega\cdot\vec\Omega'(1-L^2/2-L'^2/2)+\vec
L\cdot\vec L')\cr &=\sum(-\vec\Omega\cdot\vec\Omega'+\vec L^2+\vec
L\cdot\vec L')\,,}
$$
$$\eqalign{(\vec\Omega\wedge\vec L)\cdot(\vec\Omega'\wedge\vec L')
&=(\vec L\cdot\vec L')(\vec\Omega\cdot\vec\Omega')-
(\vec\Omega'\cdot\vec L)(\vec\Omega\cdot\vec L')\cr &\approx(\vec
L\cdot\vec L')(\vec\Omega\cdot\vec\Omega)- (\vec\Omega\cdot\vec
L)(\vec\Omega'\cdot\vec L')=(\vec L\cdot\vec L')\,,}
$$
since $\sum(\vec\Omega\cdot\vec L'-\vec\Omega'\cdot\vec L)=0.$
\medskip

\noindent {\bf Alternative formulation}
\medskip
$$z=\rho\e^{i\phi},\quad p=2S{\rho^2\over 1+\rho^2},\quad$$
$$
\langle\partial_t\rangle=ip\dot\phi.$$

$$
{\langle z|J_+|z\rangle\over\langle
z|z\rangle}=\sqrt{p(2S-p)}\e^{i\phi}, \quad {\langle
z|J_0|z\rangle\over\langle z|z\rangle}=-S+p
$$

$$
{\langle z|J_x|z\rangle\over\langle
z|z\rangle}=\sqrt{S^2-(p-S)^2}\cos({\phi})
$$
$$
{\langle z|J_y|z\rangle\over\langle
z|z\rangle}=\sqrt{S^2-(p-S)^2}\sin({\phi})
$$
$$
{\langle z|J_z|z\rangle\over\langle z|z\rangle}=-S+p
$$
Sites $\a+\h$ are nearest neighbours to $\a.$ Poisson brackets:
$\{\phi,p\} =1,\quad \{J_x,J_y\}=J_z,\quad\cdots$
Stationary phase aproximation
$${\cal L}=i\sum_\a p_\a\dot\phi_\a-H(\phi_\a,p_\a)
$$
$$
H=\sum_\a\sum_\h {\bf J}_\a\cdot{\bf J}_{\a+\h}
$$
\medskip
\noindent {\bf Formulation manifestly rotational invariant }
\medskip
$${\langle z'|\vec S| z\rangle\over\langle z'|
z\rangle}={\vec\Omega+\vec\Omega'+i[\vec\Omega\times\vec\Omega']
\over 1+\vec\Omega\cdot\vec\Omega'}
$$
Alternative kinetic energy
$${\cal T}=-2iS {\dot{\vec \Omega}\cdot\vec M\over1+\vec M^2/4},\quad\vec
M=\vec\Omega\times\vec L
$$
$${\vec n_a+\vec n_b\over 2}={\vec L\over 1+L^2/4},\quad
{\vec n_a-\vec n_b\over 2}=\vec\Omega{1-L^2/4\over 1+L^2/4},
$$
$${1-\vec n_a\cdot\vec n_b\over 2}=\left(1-L^2/4\over 1+L^2/4\right)^2
$$
$$\eqalign{{\vec L\cdot(\vec\Omega\times\dot{\vec\Omega})\over 1+L^2/4}
=&{\vec n_a+\vec n_b\over 2}\cdot(\vec
\Omega\times\dot{\vec\Omega})\cr =&{(\vec n_a+\vec n_b)\cdot[(\vec
n_a-\vec n_b) \times(\dot{\vec n_a}-\dot{\vec n_b})]\over4(1-\vec
n_a\cdot\vec n_b)}\cr =&{\vec n_b\cdot(\vec n_a\times\dot{\vec
n_a})+ \vec n_a\cdot(\vec n_b\times\dot{\vec n_b})\over 1-\vec
n_a\cdot\vec n_b} }$$
$$\e^{-iJ_y\theta}\e^{-iJ_z\phi}(J_x-iJ_y)\e^{iJ_z\phi}\e^{iJ_y\theta}=
\e^{-i\phi}{\cos\theta-1\over 2}J_++\e^{i\phi}{\cos\theta+1\over
2}J_- -\sin\theta J_z
$$
$$\e^{-iJ_y\theta}\e^{-iJ_z\phi}J_z\e^{iJ_z\phi}\e^{iJ_y\theta}=
{1\over 2}\sin\theta(\e^{-i\phi}J_++\e^{i\phi}J_-) +\cos\theta J_z
$$
If $J_-|0\rangle=0,$ then
$\e^{-iJ_y\theta}\e^{-iJ_z\phi}J_-\e^{iJ_z\phi}\e^{iJ_y\theta}|\theta,\phi
\rangle=0,$ where
$|\theta,\phi\rangle=\e^{-iJ_y\theta}\e^{-iJ_z\phi} |0\rangle.$ We
wish to find $|z\rangle$ such that
$\e^{-iJ_y\theta}\e^{-iJ_z\phi}J_-\e^{iJ_z\phi}\e^{iJ_y\theta}|z
\rangle=0.$ Clearly,
$$\eqalign{
&\e^{-zJ_+}[\e^{-iJ_y\theta}\e^{-iJ_z\phi}J_-\e^{iJ_z\phi}\e^{iJ_y\theta}]
\e^{zJ_+}\cr
&=\e^{-i\phi}{\cos\theta-1\over2}J_++\e^{i\phi}{\cos\theta-1\over2}
(J_--2zJ_z-z^2J_+)-\sin\theta(J_z+zJ_+)}
$$
Therefore
$$\e^{-i\phi}{\cos\theta-1\over2}-\e^{i\phi}{\cos\theta+1\over2}z^2-
z\sin\theta=0,\quad 2z\e^{i\phi}{\cos\theta+1\over2}+\sin\theta=0.
$$
So that
$$z=-\tan({\theta/2})\,\e^{-i\phi}.
$$
\bigskip

\end
(SUPRIMIR AT\'E \%\%\%\%\%)

{\bf Introduction}

\medskip
 Using a path integral approach over $su(2)$ coherent states
 a simple
system of two interacting spins is studied on the basis of a
rotationally invariant form of the Lagrangian recently proposed
for the description of the disordered phase of the Heisenberg
antiferromagnet.
\bigskip

\noindent {\bf Path integral over $su(2)$ coherent states}
\medskip
It is convenient to describe systems of spins in terms of $su(2)$
coherent states $|z\rangle$ defined as
$$|z\rangle=\e^{z S_+}|s,-s\rangle,$$
where $|s,-s\rangle$ is such that
$$S_-|s,-s\rangle=0, \quad
S_z|s,-s\rangle=-s|s,-s\rangle.$$ The dynamics of spin states may
be formulated in the framework of path-integrals over coherent
$su(2)$ states. The following overlaps are easily obtained
$$
\langle z_{k+1}|z_k\rangle=(1+z_{k+1}^*z_k)^{2S} \approx\langle
z_{k+1}|z_{k+1}\rangle {\rm exp}\left[-2\epsilon S{ z_{k+1}^*\dot
z_k \over 1+z_{k+1}^*z_{k+1}}\right],
$$
$$
\epsilon\dot z_k=z_{k+1}-z_k,
$$
$$\eqalign{
\langle z_{k+1}|\e^{-\epsilon{\cal H}}|z_k\rangle
&\approx\langle z_{k+1}|z_{k+1}\rangle {\rm
exp}\left[-\epsilon\left(2s{ z_{k+1}^*\dot z_k \over
1+z_{k+1}^*z_{k+1}}+{\langle z_{k+1}|{\cal H}|z_k\rangle\over
\langle z_{k+1}|z_k\rangle}\right)\right],}
$$
where ${\cal H}$ denotes the Hamiltonian operator.
We introduce the unit vector $\vec n$ parallel to the spin $\vec
S.$ We easily find
$$\eqalign{&
sn_x={\langle z|S_x|z\rangle\over\langle z|z\rangle}={2s}
{x\over1+x^2+y^2},\quad sn_y={\langle z|S_y|z\rangle\over\langle
z|z\rangle}={2s} {-y\over1+x^2+y^2},\quad\cr & sn_z={\langle
z|S_z|z\rangle\over\langle z|z\rangle}=s {x^2+y^2-1\over
1+x^2+y^2},\quad i{\langle z|\partial_t|z\rangle\over\langle
z|z\rangle}= 2s{y\dot x-x\dot y\over 1+x^2+y^2}.}$$ We also obtain
$${n_x\dot n_y-n_y\dot n_x\over2(1-n_z)}
=2{y\dot x-x\dot y\over(1+x^2+y^2)^2 (1-n_z)}={y\dot x-x\dot
y\over1+x^2+y^2}.
$$
\medskip
\bigskip
\noindent {\bf Overlap}

\%\%\%\%\%\%
This circumstance is not convenient for
subsequent nonlinear changes of variables. One can introduce new
variables ${\bf n}_{a,b}(\tau,l)$ which are both defined at the
sublattice ${\rm a}$ (or at the center of the AF elementary cell).

For that we pass to the Fourier image ${\bf n}_{a,b}(\tau,{\bf
k})$ of the original vectors ${\bf n}_a(\tau,l)$ and ${\bf
n}_b(\tau,l')$, where the momentum  vector ${\bf k}$ runs over the
AF Brillouin band. We can return to the space representation and
consider the coordinate ${\bf\rho}$ as continuous variable. As a
result we have the following definition:
$$ 
{\bf n}_{a,b}(\tau,{\bf \rho}) = \sqrt{1/N_s}\sum_{{\bf
k}}\exp\left(i{\bf k}\cdot {\bf \rho}\right) {\bf
n}_{a,b}(\tau,{\bf k}),
$$ 
where $2N_s$ is the total number of sites in the space lattice. Of
course, we assume periodic boundary conditions. We can put the
variable ${\bf \rho}$ now on the sublattice ${\rm a}$ (or in the
center of the AF elementary cell). One can check that the
Lagrangian density ${\cal L}_{kin}$ will be the same in terms of
the new variables ${\bf n}_{a,b}(\tau,{\bf\rho})\equiv{\bf
n}_{a,b}(\tau,l)$.
In the same manner one can change the measure of integration (2) 
and write out it in terms of ${\bf n}_{a,b}(\tau,l)$. The
Hamiltonian ${\cal H}$ preserves its simple form in the momentum
representation.